\documentstyle[preprint,aps,epsfig]{revtex}

\newcommand{\beq}{\begin{eqnarray}}
\newcommand{\eeq}{\end{eqnarray}}
\newcommand{\beqs}{\begin{eqnarray}}
\newcommand{\eeqs}{\end{eqnarray}}
\newcommand{\bary}{\begin{array}}
\newcommand{\eary}{\end{array}}

\newcommand{\ps}{\mbox{ps}}

\newcommand{\BR}{\mbox{BR}}

\newcommand{\figpos}{p}        
\newcommand{\spaAR}{6.5}       
\newcommand{\heightAR}{15.2}   
\newcommand{\widthAR}{15.2}    

\newcommand{\Op}{{\cal{O}}}

\newcommand{\andthis}{~~~~~\mbox{and}~~~~~}

\def\Re{{\rm Re}}
\def\Im{{\rm Im}}

\def\npb#1{Nucl.\ Phys.\ {\bf B\,#1}}

\def\plb#1{Phys.\ Lett.\ {\bf B\,#1}}

\def\prd#1{Phys.\ Rev.\ {\bf D\,#1}}
\def\prl#1{Phys.\ Rev.\ Lett. {\bf#1}}

\def\epjc#1{Eur.~Phys.~J.\ {\bf C\,#1}}

\def\progtp#1{Prog.\ Th.\ Phys.\ {\bf #1}}

\def\gsim{\ \rlap{\raise 3pt \hbox{$>$}}{\lower 3pt \hbox{$\sim$}}\ }
\def\lsim{\ \rlap{\raise 3pt \hbox{$<$}}{\lower 3pt \hbox{$\sim$}}\ }

\def\putFig#1#2#3#4#5#6#7 
{
 \newpage
 \begin{figure}[\figpos]
 \LARGE
 $#7$
 \begin{center}
 \mbox{\epsfig{figure=#1.eps,angle=0,width=#3cm,height=#4cm}}
 $#6$
 \end{center}
 \vspace{1.5cm}
 \caption{#2}
 \label{#1}
 \end{figure}
}

\begin{document}

\preprint{\vbox{\hbox{WIS-00/11/JUL-DPP}
                \hbox{hep-ph/0007170}}}

\title{~\\Standard Model correlations between 
       $K \to \pi \nu \bar\nu$ decays and observables in $B$ Physics} 

\author{Sven Bergmann and Gilad Perez \\ 
        \small \it Department of Particle Physics,
        Weizmann Institute of Science,
        Rehovot 76100, Israel}
\maketitle

\begin{abstract}
  Including the recent preliminary results of BaBar and BELLE
  experiments, we update the currently allowed intervals for various
  CKM parameters: $\bar\rho$, $\bar\eta$, $\sin2\beta$, $\sin2\alpha$,
  $\sin^2\gamma$. We also update the SM prediction for the rates of
  the $K_L \to \pi^0 \nu \bar\nu$ and $K^+ \to \pi^+ \nu \bar\nu$
  decays, their ratio $a_{\pi \nu \bar\nu}$, as well as for certain
  observables related to $B$ Physics like the CP asymmetries $a_{\psi
    K_S}$ in $B^0_d \to J/\psi \, K_S$ and $a_{\psi \phi}$ in $B^0_s
  \to J/\psi \, \phi$ or the mass differences $\Delta m_q$ ($q=d,s$)
  in the $B^0_q-\bar B^0_q$ systems. We investigate the correlations
  between them. The strongest correlations are between i) $a_{\pi \nu
    \bar\nu}$ and $a_{\psi K_S}$, ii) $\BR(K^+ \to \pi^+ \nu \bar\nu)$
  and $\Delta m_d/\Delta m_s$ and iii) $\BR(K_L \to \pi^0 \nu
  \bar\nu)$ and $a_{\psi \phi}$. These correlations are likely to be
  violated in the presence of New Physics and therefore provide
  stringent tests of the Standard Model.
\end{abstract}

\vspace{1cm}

\section{Introduction}

The CKM mixing matrix of the Standard Model (SM) provides a consistent
explanation of all experimental data on quark flavor mixing and CP
violation. Yet, some of its parameters have not been measured very
accurately.  It could well be that future experiments reveal
inconsistencies and require contributions to flavor physics from New
Physics.  In particular, among the four Wolfenstein parameters
$(\lambda,A,\rho,\eta)$~\cite{WOLF} that describe the CKM
matrix~\cite{CKM} only two have been determined with a good
accuracy~\cite{PDG}:
\beq \label{lambdaA}
\lambda = 0.2205 \pm 0.0018~~~~~\mbox{and}~~~~~A = 0.826 \pm 0.041 \,,
\eeq
while the uncertainty for $(\rho,\eta)$, which describe the apex of
the unitarity triangle from
\beq \label{unitarity}
V_{ud} V_{ub}^* + V_{cd} V_{cb}^* + V_{td} V_{tb}^* = 0 \,,
\eeq
is rather large. As a consequence the single CP violating phase of
the SM is relatively weakly determined to be:
\beq \label{delta}
\delta = \arctan(\eta/\rho) \simeq \pi/4 - \pi/2 \,.
\eeq
 
At present the only unambiguous indication for CP violation (CPV)
stems from the neutral kaon system. In particular for a long time
there has been evidence for CPV in mixing from the measurement of the
parameter~\cite{Buras}:
\beq \label{epsK}
\epsilon_K = (2.280 \pm 0.013) \times 10^{-3} \exp(i \phi_\epsilon) \,, \,
~~~~~\phi_\epsilon \simeq \pi/4 \,.
\eeq
CPV is still to be seen for other mesons, the best candidate being the
observation of a CP asymmetry in $B^0_d \to J/\psi \, K_S$ decays,
where at present $1.75\sigma$ is the combined result~\cite{ENP} of the
three most recent related experiments, CDF BaBar and
BELLE~\cite{CDF,BaBarexp,Belle}.  Another interesting signal for CPV
is the $K_L \to \pi^0 \nu \bar\nu$ decay.

The Standard Model makes definite predictions for the rates of the $K
\to \pi \nu \bar\nu$ decays and for the observables related to $B$
physics like the mass differences $\Delta m_q$ ($q=d,s$) in the
$B^0_q-\bar B^0_q$ systems and the CP asymmetries in their decays.
Combinations of some of these observables, i.e. the ratios $\Gamma[K_L
\to \pi^0 \nu \bar\nu]/\Gamma[K^+ \to \pi^+ \nu \bar\nu]$ and $\Delta
m_d/\Delta m_s$ have rather small uncertainties and therefore present
excellent probes to test the SM.  If future measurements determine any
of these variables to lie outside the intervals predicted by the SM,
this will be a clear evidence for New Physics.  Moreover, since the SM
predictions essentially depend only on the values of $\eta$ and $\rho$
there exist strong correlations between some of the observables. Then,
even if each single observable is measured to be within its SM range,
inconsistencies could still arise if correlations between various
observables are violated.

This work is organized as follows: In section~\ref{observables} we
define the relevant set of observables and discuss their experimental
status. In section~\ref{SM} we present the SM predictions for these
quantities in terms of $\rho$ and $\eta$. Our discussion contains a
careful translation of the relevant CKM matrix elements into the
Wolfenstein parameters including corrections of order $\lambda^2$ with
respect to the leading order. In section~\ref{numerical} we update the
allowed region for $(\rho,\eta)$ in order to determine the present
intervals for each observable within the SM. Then we consider the
correlations between the various observables or their combinations. We
find particularly strong correlations between i) the ratio $a_{\pi \nu
\bar\nu}$ between the $K_L \to \pi^0 \nu \bar\nu$ and $K^+ \to \pi^+
\nu \bar\nu$ decay rates and the CP asymmetry $a_{\psi K_S}$, ii)
$\BR(K^+ \to \pi^+ \nu \bar\nu)$ and $\Delta m_d/\Delta m_s$ and iii)
$\BR(K_L \to \pi^0 \nu \bar\nu)$ and $a_{\psi \phi}$. We conclude in
section~\ref{conclusion}.

\section{Observables: Definitions and experimental status}
\label{observables}

\subsection{$K \to \pi \nu \bar\nu$ decays}

The rare semi-leptonic decays $K^+ \to \pi^+ \nu \bar \nu$ and $K_L
\to \pi^0 \nu \bar\nu$ contain valuable information about the
underlying physics relevant to these processes. Since these decays are
expected to be dominated by short distance contributions they are
subject to a rather clean theoretical interpretation.

However, due to the neutrinos in the final state the $K \to \pi \nu
\bar\nu$ decays present an experimental challenge and so far only the
branching ratio (BR) of the semi-leptonic decay $K^+ \to \pi^+ \nu
\bar\nu$ has been measured with rather large uncertainty~\cite{Adler}
\beq \label{BRexpt}
\BR(K^+ \to \pi^+ \nu \bar\nu) = 
(1.5^{+3.4}_{-1.2}) \times 10^{-10} \,.
\eeq 
The detection of the $K_L \to \pi^0 \nu \bar\nu$ decay is even more
challenging: The final state has a very difficult signature since it
contains no charged particles.  At present we only have an upper bound
on the BR~\cite{KT}:
\beq 
\BR(K_L \to \pi^0 \nu \bar\nu) < 5.9 \times 10^{-7}\ \ \  
(95\,\% \ \, {\rm CL})\,,
\eeq
which lies about four order of magnitudes above the SM prediction.
However, using isospin symmetry a model independent bound has been
derived in Ref.~\cite{GN}:
\beq
\BR(K_{L}\to \pi ^{0}\nu \bar{\nu}) < 
4.4 \, \BR(K_{L}^{+}\to \pi ^{+}\nu \bar{\nu}) \,.
\eeq
Then the measurement in eq.~(\ref{BRexpt}) implies a more restrictive
upper bound:
\beq \label{lowup}
\BR(K_L\to\pi^0\nu\bar{\nu})\lsim 6.6\times 10^{-10} \,.
\eeq
Several experiments with the sensitivity to measure this BR at the SM
level have been proposed: BNL-E926 at Brookhaven~\cite{E926}, KAMI
collaboration at Fermilab~\cite{KAMI} and KEK in Japan~\cite{KEK}.

The knowledge of the above decay rates would also allow to determine
their ratio
\beq \label{apinunu}
a_{\pi \nu \bar\nu} \equiv 
{\Gamma(K_L \to \pi^0 \nu \bar\nu) \over 
 \Gamma(K^+ \to \pi^+ \nu \bar\nu)} \,,
\eeq
which is relatively clean from the theoretical point of view.

\subsection{Neutral $B$ system}

The neutral $B$ system has already began to be studied with improved
precision by the BaBar and BELLE experiments~\cite{BaBarexp,Belle} and
will be further studied with unprecedented precision by the BaBar,
BELLE, HERA-B, CLEO-III, CDF and D0 experiments~\cite{Ee}.  So far the
mass difference $\Delta m_q \equiv m(B_q^H) - m(B_q^L)$ between the
heavy ($B_q^H$) and light ($B_q^L$) mass eigenstates has only been
measured for the $B^0_d$ system~\cite{ExpMd}:
\beq \label{dmd}
\Delta m_d = (0.472 \pm 0.017) \, \ps^{-1} \,.
\eeq
For the $B^0_s$ system, there exists only a lower bound for the mass
difference~\cite{ExpMb1}:
\beq \label{dms}
\Delta m_s  > 14.6 \, \ps^{-1} \ \ \ 
(95\,\% \ \, {\rm CL})\,.
\eeq

It is useful to consider the ratio between the above mass differences
\beq \label{RmB}
R_{\Delta m_B} \equiv {\Delta m_d \over \Delta m_s} 
 \leq 0.035 \,,
\eeq
since its theoretical prediction has rather small hadronic
uncertainties.

The ratio between the width difference $\Delta\Gamma_q \equiv
\Gamma(B_q^H) - \Gamma(B_q^L)$ and the total width $\Gamma_q$ is known
to be small for the $B^0_d-\bar B^0_d$ system~\cite{Bigi}, while it is
sizeable in the $B^0_s-\bar B^0_s$ system. A recent
measurements~\cite{ExpMb1} gives
\beq
\Delta \Gamma_s/\Gamma_s = 0.17^{+0.09}_{-0.10} \,,
\label{DGammas}
\eeq
which is consistent with the SM prediction~\cite{DeltaGammaSM}.
 
A lot of information can be gained from the time-dependent CP
asymmetry of decaying $B_q^0$ mesons ($q=d,s$):
\beq \label{ACP}
A(t)_{\rm CP}^{(q)} \equiv 
 \frac{\Gamma(\bar B^0_q(t) \to \bar f) - \Gamma(B^0_q(t) \to f)}
      {\Gamma(\bar B^0_q(t) \to \bar f) + \Gamma(B^0_q(t) \to f)} \,,
\eeq
where $B^0_q(t)$ and $\bar B^0_q(t)$ refer to meson eigenstates
that have evolved from the interaction eigenstates
\beqs
B^0_q      &=& p_{B_q} B^L_q + q_{B_q} B^H_q \,, \\ 
\bar B^0_q &=& p_{B_q} B^L_q - q_{B_q} B^H_q \,,
\eeqs
after a time $t$. Let us assume that there is no CPV in mixing, i.e.
$(q/p)_{B_q}=e^{i \phi_M^{(q)}}$. Then for decays into final CP
eigenstates [CP $|f\bigr> = \pm |f\bigr>$] the asymmetry in
eq.~(\ref{ACP}) is given by
\beq \label{ACP2}
A(t)_{\rm CP}^{(q)} =
{a_{\rm CP}^{\rm dir}(B^0_q \to f) \cos(\Delta m_q t) +
 a_{\rm CP}^{\rm ind}(B^0_q \to f) \sin(\Delta m_q t) \over
 \cosh(\Delta\Gamma_q t/2) + 
 a_{\Delta\Gamma}(B^0_q \to f) \sinh(\Delta\Gamma_q t/2) } \,.
\eeq
In eq.~(\ref{ACP2}), we have separated the ``direct'' from the 
``mixing-induced'' (due to interference between mixing and decay 
amplitudes) CP-violating contributions, which are described by
\beq
a_{\rm CP}^{\rm dir}(B^0_q \to f) \equiv 
{1 - |\lambda_f^{(q)}|^2 \over
 1 + |\lambda_f^{(q)}|^2} \andthis
a_{\rm CP}^{\rm ind}(B^0_q \to f) \equiv 
{2\Im\lambda_f^{(q)} \over
 1 + |\lambda_f^{(q)}|^2} \,,
\eeq
where $\lambda_f^{(q)} \equiv (q/p)_{B_q} \cdot A(\bar B^0_q(0) \to f)
/ A(B^0_q(0) \to f)$. Note that the observable
\beq
a_{\Delta\Gamma}(B^0_q \to f) \equiv {2\Re\lambda_f^{(q)} \over
 1 + |\lambda_f^{(q)}|^2} \,,
\eeq
is not independent of $a_{\rm CP}^{\rm dir}$ and 
$a_{\rm CP}^{\rm ind}$ due to
\beq
\left[a_{\rm CP}^{\rm dir}(B^0_q \to f) \right]^2 + 
\left[a_{\rm CP}^{\rm ind}(B^0_q \to f)\right]^2 +
\left[a_\Delta(B^0_q \to f)\right]^2 = 1 \,.
\eeq

A particularly promising decay mode is the $B^0_d \to J/\psi \, K_S$
decay. There exist already results suggesting non-vanishing CP
asymmetry in this decay~\cite{CDF,BaBarexp,Belle,OPAL,ALEPH}.  Fitting
the recent experimental data~\cite{CDF,BaBarexp,Belle} to the function
in eq.~(\ref{ACP2}) in the limit where the width difference
$\Delta\Gamma_d \ll \Gamma_d$ and the asymmetry $a_{\rm CP}^{\rm
  dir}(B^0_d \to J/\psi \, K_S) \ll 1$ yields~\cite{ENP}
\beq \label{apsiKs}
a_{\psi K_S} \equiv a_{\rm CP}^{\rm ind}(B^0_d \to J/\psi \, K_S) = 
0.42 \pm 0.24 \,.
\eeq
The error of the above measurement is expected be reduced
significantly in the near future.

Finally, we turn to the $B^0_s$ decays.  The $B^0_s \to J/\psi \,
\phi$ has a simple signature and a rather large branching
fraction~\cite{PDG}, $\BR(B^0_s \to J/\psi \, \phi) = (9.3 \pm 3.3)
\times 10^{-4}$. A complete analysis of this decay appears feasible at
the LHCb, BTeV, ATLAS and CMS~\cite{Ee} because of the large
statistics and good proper time resolution of the experiments.  For
decays into two vector mesons, such as $B^0_s \to J/\psi \, \phi$, it
is convenient to introduce linear polarization amplitudes $A_0(t)$,
$A_\parallel(t)$ and $A_\perp(t)$. $A_\perp(t)$ describes a CP-odd
final-state configuration, while $A_0(t)$ and $A_\parallel(t)$
correspond to CP-even final-state configurations. In order to
disentangle them, one has to study angular distributions of the decay
products of the decay chain $B^0_s \to J/\psi [\to l^+l^-] \, \phi[\to
K^+K^-]$ (see Ref.~\cite{Dighe} for details).  Recently preliminary
results for the polarization amplitudes have been reported by the CDF
collaboration~\cite{Schmidt}:
\beqs \label{polAmps}
A_0           &=& 0.778 \pm 0.090 \pm 0.012 \,, \\
|A_\parallel| &=& 0.407 \pm 0.232 \pm 0.034 \,, \\
|A_\perp|     &=& 0.478 \pm 0.202 \pm 0.040 \,.
\eeqs

Within the approximation that $\Delta\Gamma_s \ll \Gamma_s$ and
that we can neglect $a_{\rm CP}^{\rm dir}(B^0_s \to J/\psi \, \phi)$,
the CP asymmetry in eq.~(\ref{ACP2}) reduces to~\cite{Dighe}
\beq \label{ACP4}
A(t)_{\rm CP}^{(s)} = {\cal D} \cdot
a_{\psi \, \phi} \sin(\Delta m_s t) \,. 
\eeq
Here ${\cal D}$ denotes the ``dilution'' factor given by
\beq \label{dilution}
{\cal D} = {1-D \over 1+D}~~~~~\rm{with}~~~~~
D \equiv {|A_\perp(0)|^2 \over |A_0(0)|^2 + |A_\parallel(0)|^2} \,,
\eeq
and we have introduced the abbreviation $a_{\psi \phi} \equiv a_{\rm
CP}^{\rm ind}(B^0_s \to J/\psi \, \phi)$.  The recent measurement
in~(\ref{polAmps}) implies that $D=0.3 \pm 0.4$, consistent with
theoretical estimates~\cite{BallBrown} ($D \sim 0.1-0.5$), but
suffering from rather large uncertainties. We stress that the
uncertainty in the dilution factor ${\cal D}$ is the main obstacle for
the extraction of the CP asymmetry $a_{\psi \phi}$.

The $B^0_s$ decays into a final CP eigenstate $|f\rangle$, such as
$B^0_s \to D_s^+ D_s^-$ or $J/\psi \, \eta^{(')}$, have the advantage
that the CP asymmetry is not ``diluted'' (${\cal D} =1$), which gives
rise to a cleaner measurement of the relevant parameters. However, due
to the small BR these modes have not been observed so far and their
full measurement will only become possible with the second generation
$B$ physics experiments.

\section{Standard Model picture}
\label{SM}

We shall discuss now the quantities introduced in the previous section
within the framework of the SM. We express the various observables in
terms of the {\it extended} Wolfenstein
parameters~\cite{BurasLautenberg} $\bar\rho$ and $\bar\eta$ (that
contain the most significant uncertainties) and the well-known
parameters $\lambda$ and $A$ [that are determined with good accuracy,
see eq.~(\ref{lambdaA})]. To this end we use the following expressions
for the relevant product of the CKM matrix elements
$\lambda_i=V^*_{is}V_{id}$ ($i=c,t$):
\beqs \label{ImRel}
\Re(\lambda_c) &=& -\lambda+\frac{1}{2}\lambda^3
  +\Op\left(\lambda^5\right) \,, 
\\
\Re(\lambda_t) &=& 
   \lambda^5 A^2(-1+\bar{\rho})+{1\over2}\lambda^7 A^2
   \left[1-3\bar\rho+2(\bar\rho^2+\bar\eta^2)\right]
  +\Op\left(\lambda^9\right) \,, 
\\
\Im(\lambda_t) &=& 
   \lambda^5 A^2\bar{\eta}\left(1+{1\over2}\lambda^2\right)
  +\Op\left(\lambda^9\right) \,,
\label{ImRel2}
\eeqs
which include the ${\cal O}(\lambda^2)$ corrections
to the leading result.  

\subsection{$K \to \pi \nu \bar\nu$ decays}

In the SM the BR of $K^+ \to \pi^+ \nu \bar\nu$ is predicted to
be~\cite{Buras}:
\beqs \label{bkpn}
\BR(K^+ \to \pi^+ \nu \bar\nu) &=&
 \kappa_+ \cdot {\cal B}_+(\lambda_c,\lambda_t) \,, \\
{\cal B}_+(\lambda_c,\lambda_t) &\equiv&
 \left[{\Im\lambda_t \over \lambda^5} X(x_t) \right]^2 +
 \left[{\Re\lambda_c \over \lambda} P_0(X) +
       {\Re\lambda_t \over \lambda^5} X(x_t) \right]^2 \,.
\nonumber
\eeqs
In the above expression we have factored out the constant
\beq \label{kappa}
\kappa_+ \equiv {3\alpha^2 \, r(K^+) \BR(K^+ \to \pi^0 e^+ \nu) \over 
 2\pi^2 \sin^4 \theta_{\rm W}} \cdot \lambda^8 = 4.11 \times 10^{-11} \,,
\eeq
including all well-determined parameters, i.e.
\beq 
\alpha                 = \frac{1}{129} \,, \qquad 
\sin^2\theta_{\rm W}   = 0.23 \,, \qquad
\BR(K^+\to\pi^0e^+\nu) = 4.82 \times 10^{-2} \,
\eeq
and $r(K^+)=0.901$~\cite{ParsaMariano}, which summarizes the isospin
breaking correction when expressing the hadronic matrix element of
$K^+ \to \pi^+ \nu \bar\nu$ in terms of the one for $K^+ \to \pi^0 e^+
\nu$.  The function ${\cal B}_+(\lambda_c,\lambda_t)$ contains the
less known parameters of the theory. $X(x_t)$ and $P_0(X)$ represent
the NLO electroweak loop contributions associated with intermediate
top and charm quarks, respectively.  We want to express the BR in
eq.~(\ref{bkpn}) as a function of $\bar\rho$ and $\bar\eta$.  Using
eqs.~(\ref{ImRel})--(\ref{ImRel2}) we find that
\beq \label{Bplus}
{\cal B}_+(\bar\rho,\bar\eta) = X^2(x_t) A^4
\left\{ \bar\eta^2 +\left[\bar\rho - (1+\Delta) \right]^2 + 
        \lambda^2 f(\bar\rho,\bar\eta) \right\} + \Op(\lambda^4) \,,
\eeq
where
\beq
f(\bar\rho,\bar\eta) \equiv 
 \bar\eta^2 + (\bar\rho-1-\Delta)(1-3\bar\rho+2\bar\rho^2+2\bar\eta^2+\Delta)
\eeq
contains the $\lambda^2$ corrections and $\Delta \equiv {P_0(X) \over
A^2 X(x_t)}$.

Let us turn now to the neutral kaon decay, $K_L \to \pi^0 \nu
\bar\nu$.  In the SM its BR is predicted to be~\cite{Buras}:
\beqs \label{BRKLSM}
\BR(K_L \to \pi^0 \nu \bar\nu) &=& 
 \kappa_L \cdot {\cal B}_L(\lambda_t) \,, \\
{\cal B}_L(\lambda_t) &\equiv&
 \left[{\Im\lambda_t \over \lambda^5} X(x_t) \right]^2 \,.
\nonumber
\eeqs
In the above expression we have factored out the constant 
\beq \label{kappaL}
\kappa_L \equiv {\tau(K_{\rm L}) \over \tau(K^+)} \cdot 
 {3\alpha^2 \, r(K_L) \BR(K^+ \to \pi^0 e^+ \nu)\over 2\pi^2
 \sin^4 \theta_{\rm W}} \, \lambda^8 = 
 1.80 \times 10^{-10} \,,
\eeq
where $r(K_L)=0.944$~\cite{ParsaMariano} contains the isospin breaking
correction when expressing the hadronic matrix element of $K_L \to
\pi^0 \nu \bar\nu$ in terms of the one for $K^+ \to \pi^0 e^+ \nu$.
We note that the leading CP violating effect for the $K^0 \to \pi^0
\nu \bar\nu$ decay arises from interference between mixing and decay,
i.e. $\Im\lambda_K \ne 0$, where $\lambda_K \equiv (q/p)_K \cdot (\bar
A/A)_K$, while contributions to CPV in mixing ($|q/p| \neq 1$) and
decay ($|\bar A /A| \neq 1$) are of order ${\cal O}(10^{-3})$ and
therefore negligible. Also note that the phase from $(q/p)_K$ is of
order $\Op\left(\lambda^4\right)$ in our parameterization and
therefore suppressed in eq.~(\ref{BRKLSM}).

The function ${\cal B}_L(\lambda_t)$ contains the less known parameter
of the theory, i.e.  $\Im\lambda_t$.  Note that to an excellent
approximation~\cite{IL} $K_L \to \pi^0 \nu \bar\nu$ is a purely CP
violating process~\cite{BIP}.  As a consequence its rate would vanish
in the limit of a real CKM matrix as can be seen from
eq.~(\ref{BRKLSM}).  This is also manifest when expressing ${\cal
B}_L(\lambda_t)$ in terms of the Wolfenstein parameters:
\beq \label{Blong}
{\cal B}_L(\bar\eta) = 
 X^2(x_t) A^4 \cdot \bar\eta^2 (1 + \lambda^2) + \Op(\lambda^4) \,.
\eeq
From eq.~(\ref{Blong}) it follows that $\BR(K_L \to \pi^0 \nu
\bar\nu)$ is proportional to the height squared of the unitarity
triangle, which is a direct measure of CPV.

The SM expression for the ratio $a_{\pi \nu \bar\nu}$ between the two
decay rates, defined in eq.~(\ref{apinunu}), follows immediately from
eq.~(\ref{bkpn}) and eq.~(\ref{BRKLSM}):
\beqs
a_{\pi \nu \bar\nu} &=&
 {r(K_L) \over r(K^+)} \cdot 
 \frac{\left({\Im\lambda_t}\right)^2}
  {\left({\Im\lambda_t}\right)^2+
   \left({\lambda^4 P_0(X) \over X(x_t)} \Re\lambda_c + \Re\lambda_t
 \right)^2} \,,
\eeqs
where we have factored out the isospin correction factor ${r(K_L)
\over r(K^+)}=1.048$~\cite{ParsaMariano}.  In terms of the Wolfenstein
parameters $\bar\rho$ and $\bar\eta$ we have~\cite{GN}
\beqs \label{Apinunu}
a_{\pi \nu \bar\nu}(\bar\rho, \bar\eta) &=& 
 {r(K_L) \over r(K^+)} \cdot \sin^2 \theta \,, \\
\sin^2 \theta &\equiv& 
  {{\cal B}_L(\bar\eta) \over {\cal B}_+(\bar\rho,\bar\eta)}  
  = {\bar\eta^2 \over \bar\eta^2 + \left[\bar\rho - (1+\Delta) \right]^2} 
 \left[1+\lambda^2 g(\bar\rho, \bar\eta) \right]
+\Op(\lambda^4) \label{Apinunu2} \,,
\eeqs
where 
\beq
g(\bar\rho, \bar\eta) \equiv 
1 - {f(\bar\rho, \bar\eta) \over 
 \bar\eta^2 + \left[\bar\rho - (1+\Delta) \right]^2}
\eeq
contains the $\lambda^2$ corrections.

\subsection{Neutral $B$ system}

The ratio $R_{\Delta m_B}$ between the mass differences $\Delta m_d$
and $\Delta m_s$, defined in eq.~(\ref{RmB}), takes the following
value in the SM~\cite{Falk,BB}:
\beqs \label{RmBSM}
R_{\Delta m_B} 
&=& \xi^{-2} {m_{B_d} \over m_{B_s}} \left|{V_{td} \over V_{ts}}\right|^2 \\
&=& \xi^{-2} {m_{B_d} \over m_{B_s}} \lambda^2
\left[\bar\eta^2 +(\bar\rho-1)^2 \right] 
\left[1+\lambda^2(1-2\bar\rho)\right]
+\Op(\lambda^4) \,,
\eeq
where the ratio of the $B_d^0$ and $B_s^0$ masses is
$m_{B_d}/m_{B_s}=0.985$~\cite{PDG} and the hadronic parameter $\xi
\equiv f_{B_s}/f_{B_d} \cdot \sqrt{B_{B_s}/B_{B_d}}$ approaches unity
in the limit of $SU(3)$ flavor symmetry. We use $\xi = 1.14 \pm
0.08$\,, as estimated from lattice
calculations~\cite{Buras,Falk,Sharpe}.

Within the SM the CP asymmetry $A(t)_{\rm CP}^{(d)}$ in the $B^0_d \to
J/\psi \, K_S$ decays, defined in eq.~(\ref{ACP}), is known to provide
an accurate measurement of the angle $\beta$ of the unitarity
triangle.  This is because $\Delta \Gamma_d \ll \Gamma_d$, such
that the denominator in eq.~(\ref{ACP2}) is unity to a very good
approximation.  Moreover, within the SM, $B^0_d \to J/\psi \, K_S$ is
dominated by the tree-level diagram while penguin contributions are
non-significant, such that one can neglect direct CP violation and set
$a_{\rm CP}^{\rm dir}(B^0_d \to J/\psi \, K_S) = 0$.  Then
eq.~(\ref{ACP2}) reduces to
\beq \label{ACP3}
A(t)_{\rm CP}^{(d)} = 
a_{\rm CP}^{\rm ind}(B^0_d \to J/\psi \, K_S) \sin(\Delta m_d \, t) \,, 
\eeq
and the measurement quoted in eq.~(\ref{apsiKs}) determines~\cite{NQ}:
\beq \label{apsiKsSM}
a_{\psi K_S} = \Im\left({V^*_{tb}V_{td}{V_{cb}V_{cd}^*} \over
                        {V_{tb}V_{td}^*} V^*_{cb}V_{cd}} \right)
= \sin 2\beta \,.
\eeq
In terms of the extended Wolfenstein parameters we have
\beq
a_{\psi K_S} \label{apsiKS}
= {2\bar\eta(1-\bar\rho) \over \bar\eta^2 + (\bar\rho-1)^2}
+ \Op\left(\lambda^4\right) \,.
\eeq 
Note that corrections to the expression in eq.~(\ref{apsiKS}) only
appear at $\Op\left(\lambda^4\right)$.

Finally, we turn to the SM prediction for the $B^0_s$ decays.  Let us
discuss first the $B_s^0$ decay into a final CP eigenstate
$|f\rangle$, such as $D_s^+ D^-_s$ or $J/\psi \, \eta^{(')}$.
According to eq.~(\ref{DGammas}) the width difference $\Delta\Gamma_s$
is likely to be sizeable such that the $a_{\Delta\Gamma}(B^0_s \to f)$
term in eq.~(\ref{ACP2}) could be significant.  Moreover, neglecting
$a_{\rm CP}^{\rm dir}(B^0_s \to f)$ is a less safe assumption for
$B^0_s$ decays than for the $B^0_d$ decays, since in the SM the
subleading penguin contributions could be as large as
10\%~\cite{Fleischer}. For simplicity we assume that the issue of how
to extract $a_{\rm CP}^{\rm ind}(B^0_s \to f)$ with sufficient
accuracy will eventually be resolved. Then it is straightforward to
use the SM prediction~\cite{Fleischer}
\beqs \label{apsiphi}
a_{\rm CP}^{\rm ind}(B^0_s \to f) &=&
\Im\left({V^*_{tb}V_{ts}  V_{cb}V_{cs}^* \over
         {V_{tb}V_{ts}^*} V^*_{cb}V_{cs}} \right)
\equiv \sin 2\beta_s \\
&=&
2\lambda^2 \bar\eta \left[1+\frac{1}{2}\lambda^2(1-2\bar\rho) \right]
 +\Op\left(\lambda^4\right)
\eeqs
in order to extract the Wolfenstein parameters $\bar\rho$ and
$\bar\eta$.

At present the most promising candidate to measure $\beta_s$ is the
$B^0_s \to J/\psi \, \phi$ decay. The most significant obstacle in
extracting $\beta_s$ from $a_{\psi \, \phi}$ is due to the uncertainty
of the``dilution'' factor ${\cal D}$ in eq.~(\ref{dilution}), but
eventually it should be possible to measure and/or predict ${\cal D}$
with sufficient accuracy. Then a calculation of the penguin
contributions would be particularly important for a theoretically
clean extraction of $\beta_s$.

\section{Numerical Analysis}
\label{numerical}

In this section we present the numerical results of our analysis.
Having expressed all the relevant quantities in terms of $\bar\rho$
and $\bar\eta$, we only need to obtain the allowed region for these
two parameters in order to predict the SM values for the observables
under study.  The procedure of how to determine the relevant parameter
space for the two Wolfenstein parameters $\bar\rho$ and $\bar\eta$ is
well-known (for details see Ref.~\cite{BaBar,Buras}). Using the
updated input parameters in Tab.~1 yields the allowed region shown in
Fig.~\ref{nARrhoeta}. It is determined by (a) the measurement of
$|V_{ub}/V_{cb}|$ (corresponding to the dashed circles), (b) the
observed $B_d^0 - \bar B_d^0$ mixing parameter $\Delta m_d$
(corresponding to the dotted circles), (c) the upper bound on the
$B_s^0 - \bar B_s^0$ mixing parameter $\Delta m_s$ (which corresponds
to the dashed-dotted circle), (d) the measurement of $\epsilon_K$
(corresponding to the solid hyperbolae) and (e) the combined result of
the CDF, BELLE and BaBar measurements of $a_{\psi K_S}$ (thick grey
lines). Note that for simplicity we naively combined the above
mentioned constraints in order to determine the allowed region, which
is sufficient for the purposes of this work. (For a more accurate
analysis using a $\chi^2$ fit, see Ref.~\cite{PS}.)

\subsection{CKM Parameters}

A scan over the presently allowed region in the $(\bar\rho,\bar\eta)$
plane (the grey area in Fig.~\ref{nARrhoeta}) yields the following
intervals for the CKM parameters:
\beqs
\bar\rho &\in& [0.004, 0.27]  \,,\nonumber \\
\bar\eta &\in& [0.26, 0.37] \,.
\label{roet}
\eeqs 
Eq.~(\ref{roet}) yields an allowed range for $\delta$, the 
CP violating phase of the SM:
\beq \label{delta1}
\delta = \arctan(\eta/\rho) \simeq 45^o  - 89^o \,.
\eeq

For completeness we also update the allowed intervals for the angles
of the unitary triangle:
\beqs
\sin2\alpha  &\in& [-0.74, 0.57]  \,, \\
\sin2\beta   &\in& [0.58, 0.66]   \,, \\
\sin^2\gamma &\in& [0.51, 1.0]    \,,
\eeqs 
with $\alpha=-{\rm arg}\left({V_{td}V_{tb}^*\over V_{ud}V_{ub}^*}\right)$
and $\gamma=\pi-\alpha-\beta$.

\subsection{Individual Observables}

The scan over the gray region in the $(\bar\rho,\bar\eta)$
plane (Fig.~\ref{nARrhoeta}) yields the following
allowed intervals:
\beqs
\BR(K^+ \to \pi^+ \nu \bar\nu) &\in& [5.43, 9.96] \times 10^{-11} \,, \\
\BR(K_L \to \pi^0 \nu \bar\nu) &\in& [0.98, 3.47] \times 10^{-11} \,, \\
a_{\pi \nu \bar\nu}            &\in& [0.040, 0.087] \,.
\eeqs 

Similarly, a scan over allowed region in the $(\bar\rho,\bar\eta)$
plane for the observables in the $B^0_q-\bar B^0_q$ systems yields
\beqs
R_{\Delta m_B} &\in& [0.020, 0.035] \,, \\
a_{\psi K_S}   &\in& [0.58, 0.66]   \,, \\
a_{\psi \phi}  &\in& [0.016, 0.056] \,. 
\eeqs 
%

\subsection{Correlations}

So far we have only determined the allowed intervals for each of the
different variables under study.  Additional constraints arise when
considering the correlations between these variables. Since, within
SM, all the observables are essentially functions of two variables, i.e.
$\bar\rho$ and $\bar\eta$, in general they are not independent from
each other. Then plotting the allowed region in the parameter space of
any pair of observables does not result into the rectangle corresponding
to the product of the individual intervals determined above, but only
to some subspace of this area.

As an example consider the correlation between the ratio $a_{\pi \nu
\bar\nu}$ defined in eq.~(\ref{apinunu}) and the parameter $a_{\psi
K_S}$ that describes the CP asymmetry in $B^0_d \to J/\psi \, K_S$
decays~\cite{BB1,BB2,NirWor}, c.f.~eqs.~(\ref{apsiKs})
and~(\ref{apsiKsSM}).  Recall the SM predictions that $a_{\psi
K_S}=\sin 2\beta$ and that $a_{\pi \nu \bar\nu} = 1.048 \sin^2
\theta$. If the contribution from the $c$ quark in eq.~(\ref{bkpn})
were negligible, i.e. $\Delta=0$, then to the lowest order in
$\lambda$, the angle $\theta$ would coincide with the angle $\beta$ of
the unitarity triangle.  However, since $\Delta \sim 0.4$, this
relation is somewhat distorted.  This can be seen in
Fig.~\ref{apini_vs_apsiK}, where the SM relation between $a_{\pi
\nu\bar\nu}$ and $a_{\psi K_S}$ is shown for the input parameters in
Tab.~1.  The solid curves displays $a_{\pi \nu \bar\nu}$ as a function
of $a_{\psi K_S}$ for $\Delta=\lambda=0$.  Only in this case there is
a one-to-one correspondence between $\sin 2\beta$ and $\sin^2 \theta =
(1-\sqrt{1-\sin^2 2\beta})/2$\,.  A scan over the presently allowed
region in the $(\bar\rho,\bar\eta)$ plane (see Fig.~\ref{nARrhoeta})
yields the dark area in the $a_{\psi K_S} - a_{\pi \nu \bar\nu}$
plane, when taking the central value of $\Delta$. This region is
``smeared'' to the light area when scanning over all possible values
for $\Delta$.  In Fig.~\ref{apini_vs_apsiK_ZOOM} the correlation
region in the $a_{\psi K_S}-a_{\pi \nu \bar\nu}$ plane is magnified.

Similarly any pair of observables that functionally depend on each
other to a good approximation, is expected to be strongly correlated.
Then it follows, that the best remaining candidates for strong
correlations are $\BR(K_L \to \pi^0 \nu \bar\nu)$ vs $a_{\psi\phi}$
and $\BR(K^+ \to \pi^+ \nu \bar\nu)$ vs $R_{\Delta m_B}$~\cite{BB}.
To leading order in $\lambda$ and in the limit where $\Delta \to 0$
both $\BR(K^+ \to \pi^+ \nu \bar\nu)$ and $R_{\Delta m_B}$ are
proportional to $\bar\eta^2 + (\bar\rho-1)^2$. Therefore it is not
surprising that there is a rather strong correlation between these two
observables as can be seen from the dark narrow band in
Fig.~\ref{BrKpl_vs_Rdelmb}\,, which corresponds to the central values
for the various parameters appearing in the prefactors for both
observables.  However varying these parameters within their allowed
region, significantly enlarges the valid parameter space in the
$R_{\Delta m_B}-\BR(K^+ \to \pi^+ \nu\bar\nu)$ plane, yielding the
light area in Fig.~\ref{BrKpl_vs_Rdelmb}\,.

To leading order in $\lambda$ both $\BR(K_L \to \pi^0 \nu \bar\nu)$
and $\sin 2\beta_s$ depend just on $\bar\eta$. Only the NLO correction
in $\lambda$ introduces some $\bar\eta$ dependence in $\sin 2\beta_s$.
As we have mentioned in section~\ref{SM} it is non-trivial to extract
the value of $\sin 2\beta_s$ from the time-dependent asymmetry $a_{\rm
CP}^{\rm ind}(B^0_s \to f)$. To be explicit, we show in
Fig.~\ref{apsiphi_vs_BrKL} the correlation between $\BR(K_L \to \pi^0
\nu \bar\nu)$ and $a_{\psi\phi}$, but allow for a large uncertainty
in the ``dilution'' factor ${\cal D} = (1-D)/(1+D)$. Indeed, for the
central value $D=0.3$ there is a strong correlation between $\BR(K_L
\to \pi^0 \nu \bar\nu)$ and $a_{\psi\phi}$ corresponding to the dark
narrow band.  However, due to the present ignorance of the precise
value of $D$ we need to scan over a rather large interval ($D \in
[0.1,0.5]$) which introduces a significant smearing resulting in the
light area in Fig.~\ref{apsiphi_vs_BrKL}\,.

Examining explicitly all the remaining pairs of the observables we
found that they are less correlated than the three pairs discussed
above. Nevertheless, combining quantities with rather small
uncertainties, i.e. $\BR(K_L \to \pi^0 \nu \bar\nu)$ vs $a_{\psi K_S}$
(Fig.~\ref{BrKL_vs_apsiK}) and $\BR(K_L \to \pi^0 \nu \bar\nu)$ vs
$R_{\Delta m_B}$ (Fig.~\ref{BrKL_vs_Rdelmb}) is still useful, since
the forbidden (white) regions in the respective parameter spaces are
sizable.

\section{Conclusions}
\label{conclusion}

We have studied a set of variables related to $K \to \pi \nu \bar\nu$
decays and observables of the $B^0_d-\bar B^0_d$ and $B^0_s-\bar
B^0_s$ systems, which have been observed already (with large
uncertainties) or will be measured soon. We have focused on the SM
predictions both for the individual intervals as well as the
correlations between these variables.  The latter significantly
improve the ability to test the SM predictions: Even if future
measurements would be consistent with the individual intervals for the
various quantities, combining two (or more measurements) can easily
reveal inconsistencies of the CKM picture, which call for New Physics.
Particularly strong correlations exist between between i) $a_{\pi \nu
\bar\nu}$ and $a_{\psi K_S}$, ii) $\BR(K^+ \to \pi^+ \nu \bar\nu)$ and
$\Delta m_B/\Delta m_{B_s}$ and iii) $\BR(K_L \to \pi^0 \nu \bar\nu)$
and $a_{\rm CP}^{\rm ind}(B^0_s \to f)$. These correlations are likely
to be violated in the presence of New Physics and therefore provide
excellent tests of the Standard Model.

\vspace{2.5cm}

\acknowledgements
 
We thank Y. Nir for helpful discussions and comments on the
manuscript.

\newpage


\vspace{2.5cm}

\begin{center} 
Tab.~1: Input values \\
\vspace{1.5cm}
\begin{tabular}{|c|c|c|} 
\hline
~Parameter~ & Value \cr
\hline \hline
$a_{\psi K_S}$   & $0.42 \pm 0.24$          \cr
\hline
$R_u$            & $0.39 \pm 0.07$          \cr
\hline
$V_{td}$         & $[7.0,9.3] \, 10^{-3}$   \cr
\hline
$V_{cb}$         & $0.040 \pm 0.002$        \cr
\hline
$\Delta m_d$     & $(0.471 \pm 0.016)$ ps$^{-1}$ \cr
\hline
$\Delta m_{B_s}$ & $> 14.6$ ps$^{-1}$       \cr
\hline
$\eta_2$         & $0.57 \pm 0.01$          \cr
\hline
$\eta_X$         & $0.994$                  \cr
\hline
$\xi$            & $1.14 \pm 0.08$          \cr
\hline
$P_0(\epsilon)$  & $0.31 \pm 0.05$          \cr
\hline
$P_0(X)$         & $0.42 \pm 0.06$          \cr
\hline
$m_t$            & $(165 \pm 5)$ GeV        \cr
\hline
$A$              & $0.826 \pm 0.041$        \cr
\hline
$\lambda$        & $0.22$                   \cr
\hline  
$\Delta$         & $0.31 - 0.54$            \cr
\hline\hline
\end{tabular}
\end{center}

{}


\putFig{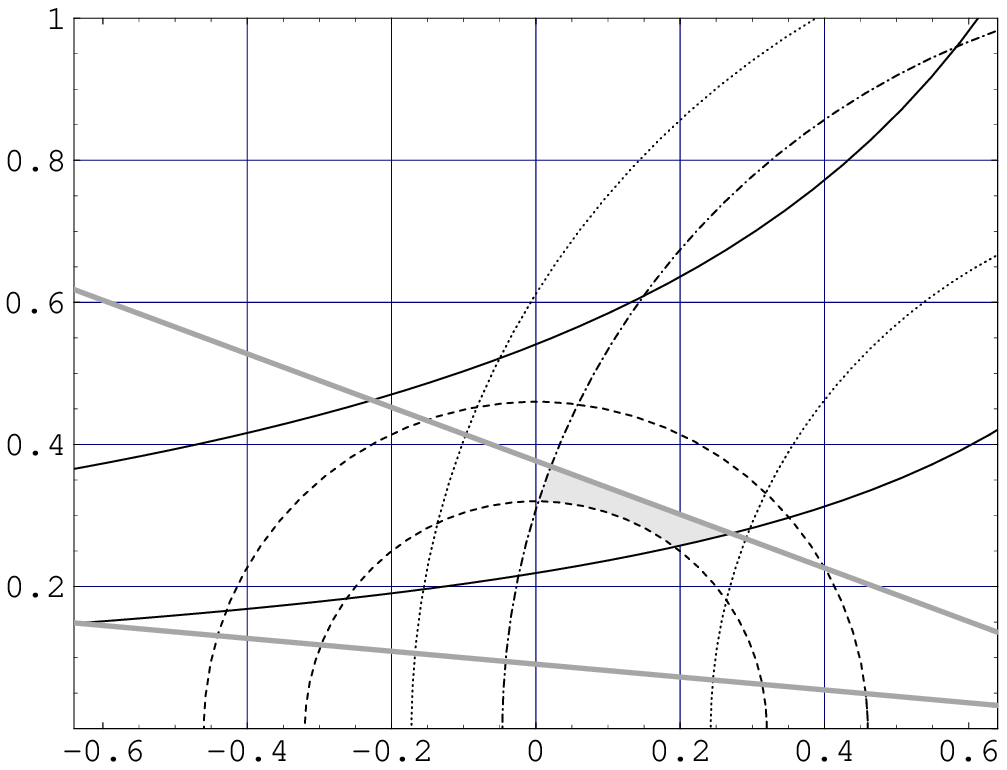}{Determination of the apex $(\bar\rho,\bar\eta)$ of
  the unitarity triangle: The grey region is the presently allowed
  region, which is determined by (a) the measurement of
  $|V_{ub}/V_{cb}|$ (corresponding to the dashed circles), (b) the
  observed $B_d^0 - \bar B_d^0$ mixing parameter $\Delta m_d$
  (corresponding to the dotted circles), (c) the upper bound on the
  $B_s^0 - \bar B_s^0$ mixing parameter $\Delta m_s$ (which
  corresponds to the dashed-dotted circle), (d) the measurement
  of $\epsilon_K$ (corresponding to the solid hyperbolae) and (e)
  the combined result of the CDF, BELLE and BaBar measurements
  of $a_{\psi K_S}$ (thick grey lines).}
  {12}{12}{8}{\bar\rho}{\bar\eta}
  
\putFig{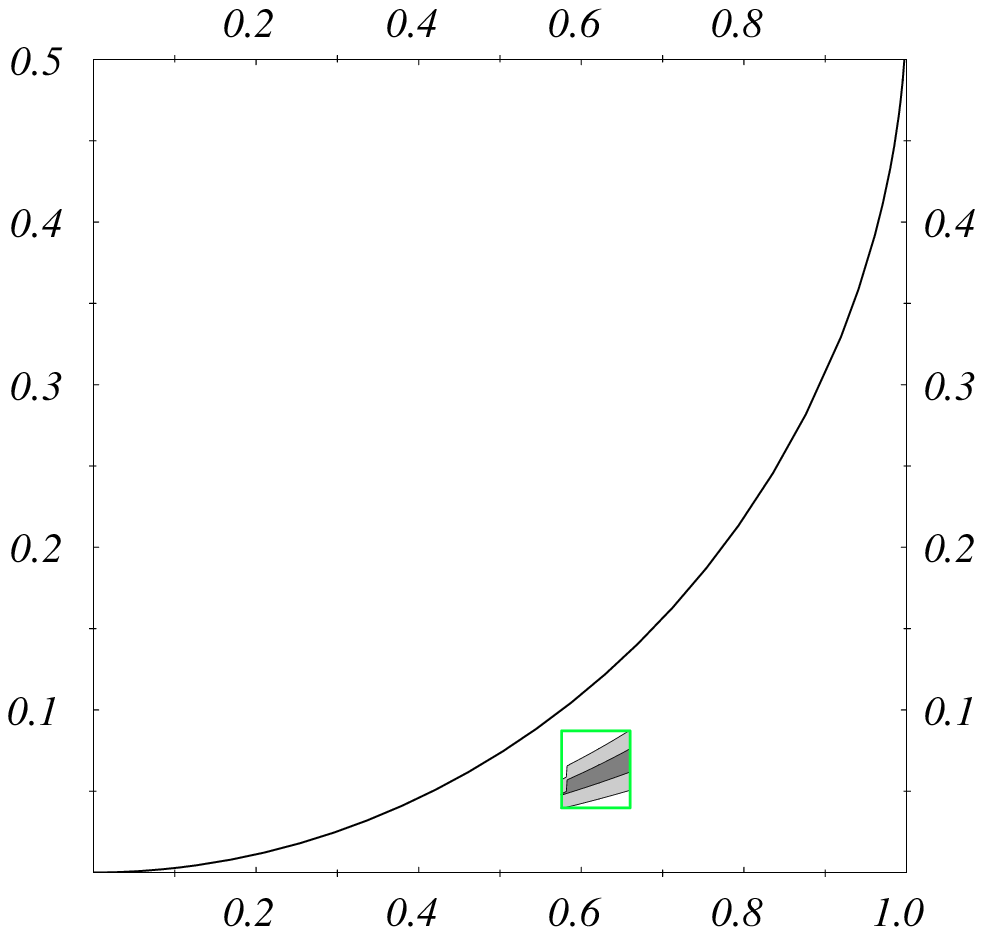}{Relation between $a_{\pi \nu \bar\nu}$ and
  $a_{\psi K_S}$ in the SM: The solid curves displays $a_{\pi \nu
  \bar\nu} (a_{\psi K_S})$ for $\Delta=0$. Only in this case there is
  a one-to-one correspondence between $a_{\pi \nu \bar\nu}$ and
  $a_{\psi K_S}$.  A scan over the presently allowed region in the
  $(\bar\rho,\bar\eta)$ plane (see Fig.~\ref{nARrhoeta}) yields the
  dark area in the $a_{\psi K_S} - a_{\pi \nu \bar\nu}$ plane, when
  taking the central value $\Delta=0.425$.  The light region
  corresponds to the interval $\Delta \in [0.31,0.54]$.}
  {\widthAR}{\heightAR}{\spaAR} {a_{\psi K_S}}{a_{\pi \nu \bar\nu}}
  
\putFig{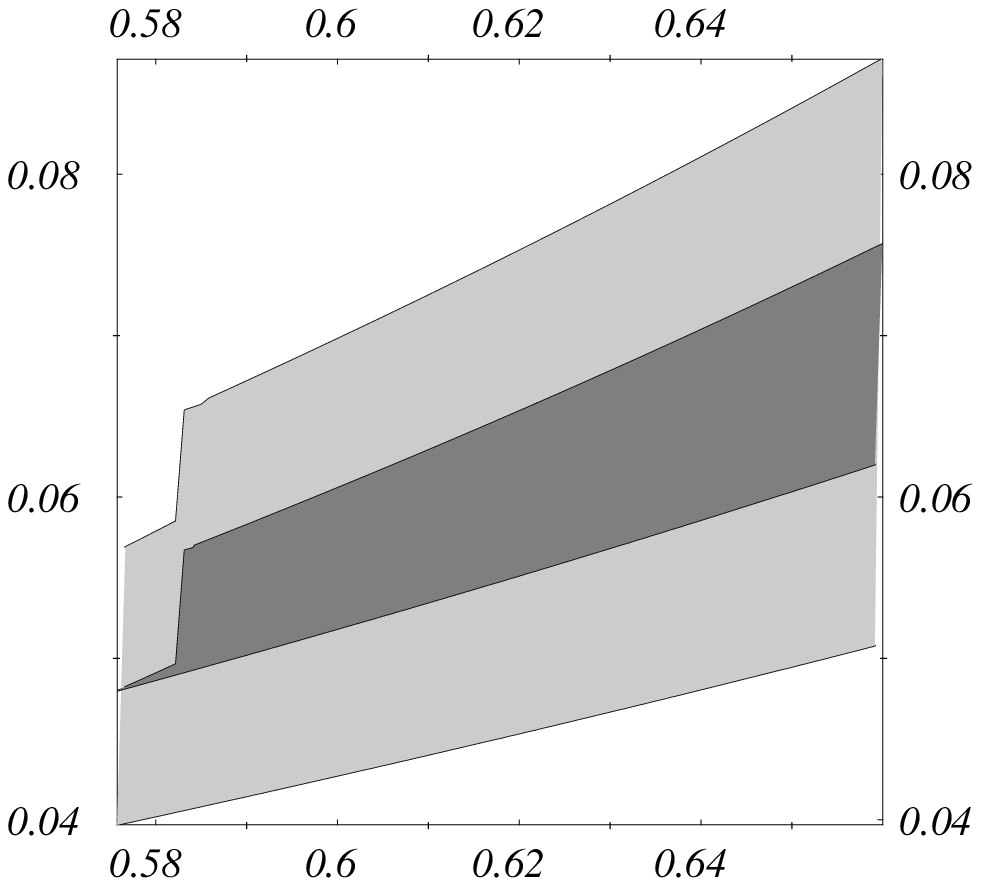}{Relation between $a_{\pi \nu \bar\nu}$
  and $a_{\psi K_S}$ in the SM.  Taking $\Delta=0.425$ yields the dark
  area in the $a_{\psi K_S}-a_{\pi \nu \bar\nu}$ plane.  The light
  region corresponds to the interval $\Delta \in [0.31,0.54]$.}
  {\widthAR}{\heightAR}{\spaAR}{a_{\psi K_S}}{a_{\pi \nu \bar\nu}}
  
\putFig{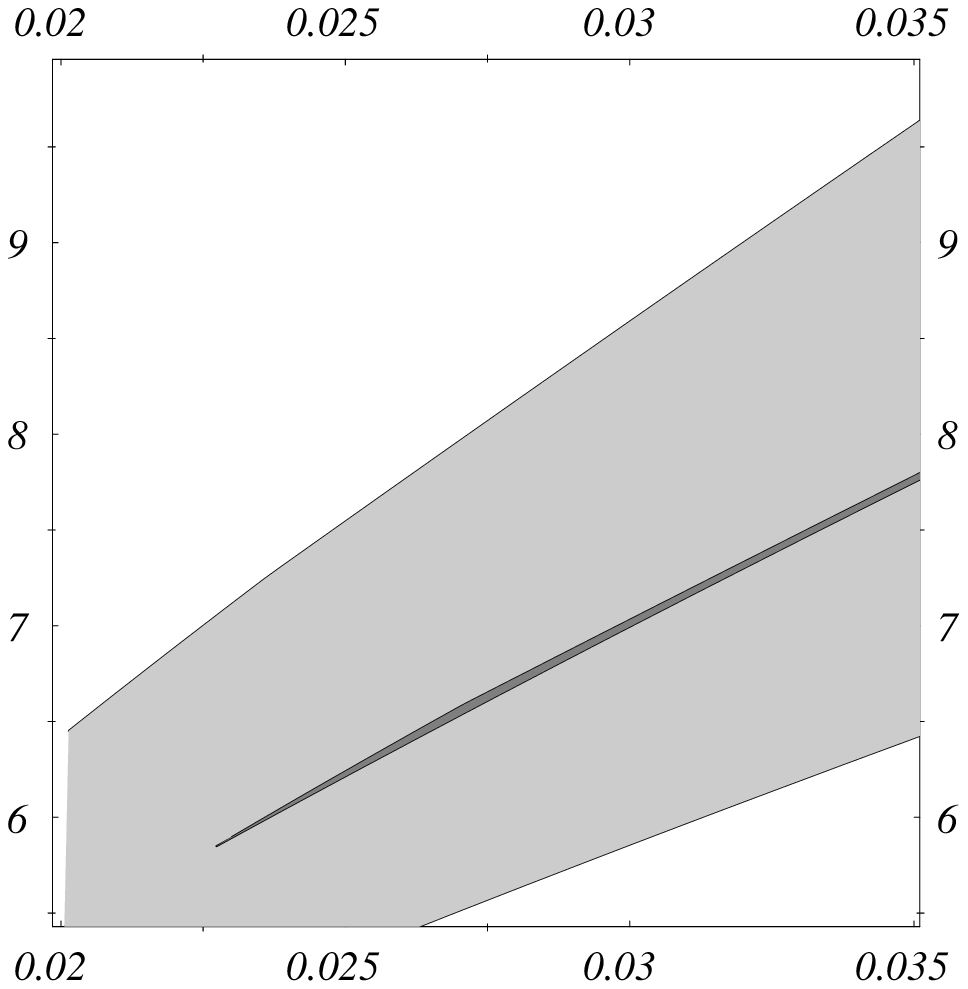}{Relation between $\BR(K^+ \to \pi^+ \nu
  \bar\nu)$ and $R_{\Delta m_B}$ in the SM: A scan over the presently
  allowed region for ($\bar\rho,\bar\eta$) (c.f. Fig.~\ref{nARrhoeta})
  yields the dark (light) region in the $R_{\Delta m_B} - \BR(K^+ \to
  \pi^+ \nu \bar\nu)$ plane for the central values (total intervals)
  of the parameters $X^2(x_t), A, \Delta$ and $\xi$.}
  {\widthAR}{\heightAR}{\spaAR}{R_{\Delta m_B}}{\BR(K^+ \to \pi^+ \nu
  \bar\nu) \times 10^{11}}

\putFig{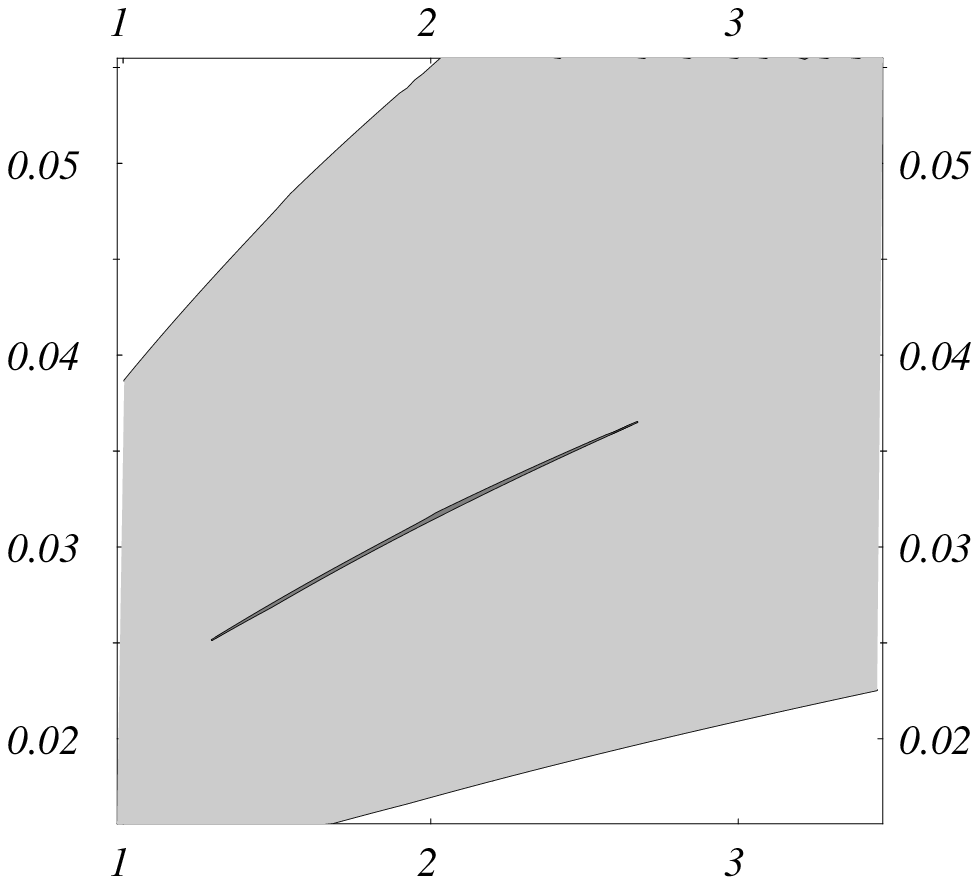}{Relation between $a_{\psi \phi}$ and $\BR(K_L
  \to \pi^0 \nu \bar\nu)$ in the SM: A scan over the presently allowed
  region for ($\bar\rho,\bar\eta)$ (c.f. Fig.~\ref{nARrhoeta}) yields
  the dark (light) region in the $\BR(K_L \to \pi^0 \nu \bar\nu) -
  a_{\psi \phi}$ plane for the central values (total intervals) of the
  parameters $X^2(x_t), A, \xi$ and ${\cal D}$.}
  {\widthAR}{\heightAR}{\spaAR}{\BR(K_L \to \pi^0 \nu \bar\nu) \times
  10^{11}}{a_{\psi \phi}}

\putFig{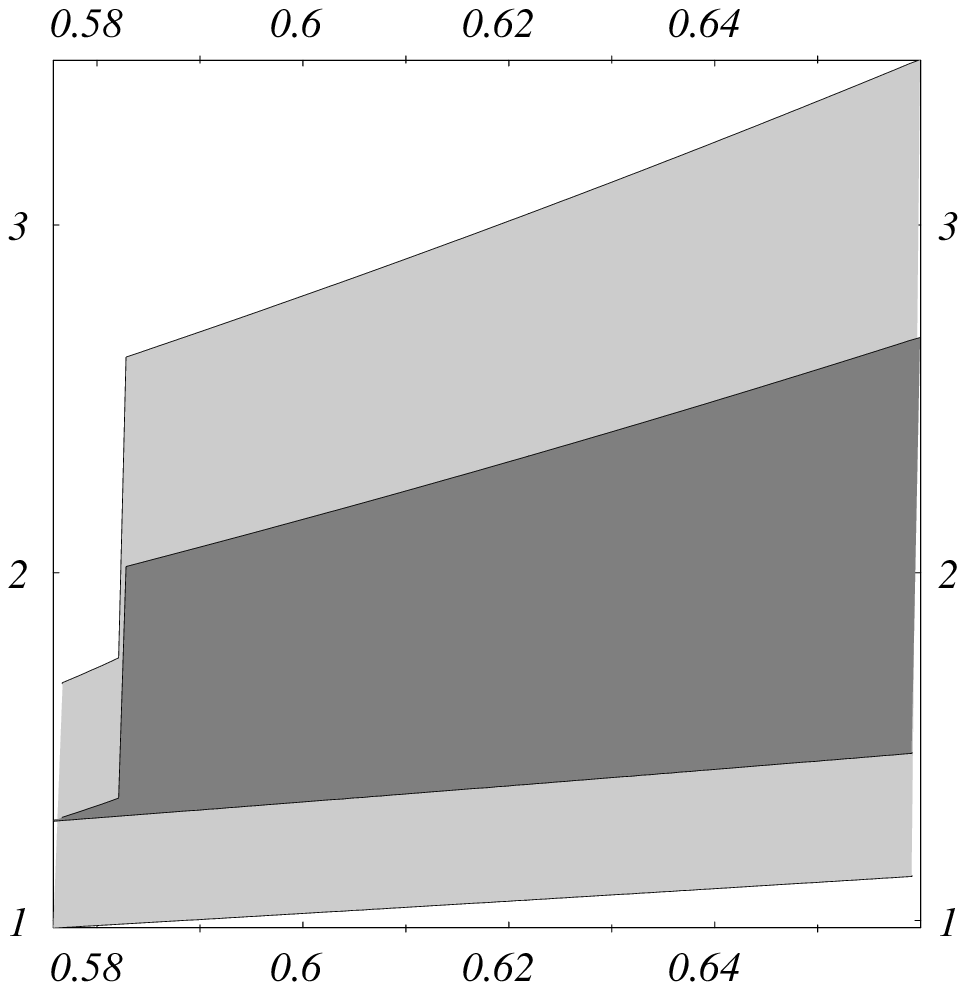}{Relation between $\BR(K_L \to \pi^0 \nu
  \bar\nu)$ and $a_{\psi K_S}$ in the SM: A scan over the presently
  allowed region for ($\bar\rho,\bar\eta$) (c.f. Fig.~\ref{nARrhoeta})
  yields the dark (light) region in the $a_{\psi K_S} - \BR(K_L \to
  \pi^0 \nu \bar\nu)$ plane for the central values (total intervals)
  of the parameters $X^2(x_t)$ and $A$.} {\widthAR}{\heightAR}{\spaAR}
  {a_{\psi K_S}}{\BR(K_L \to \pi^0 \nu \bar\nu) \times 10^{11}}

\putFig{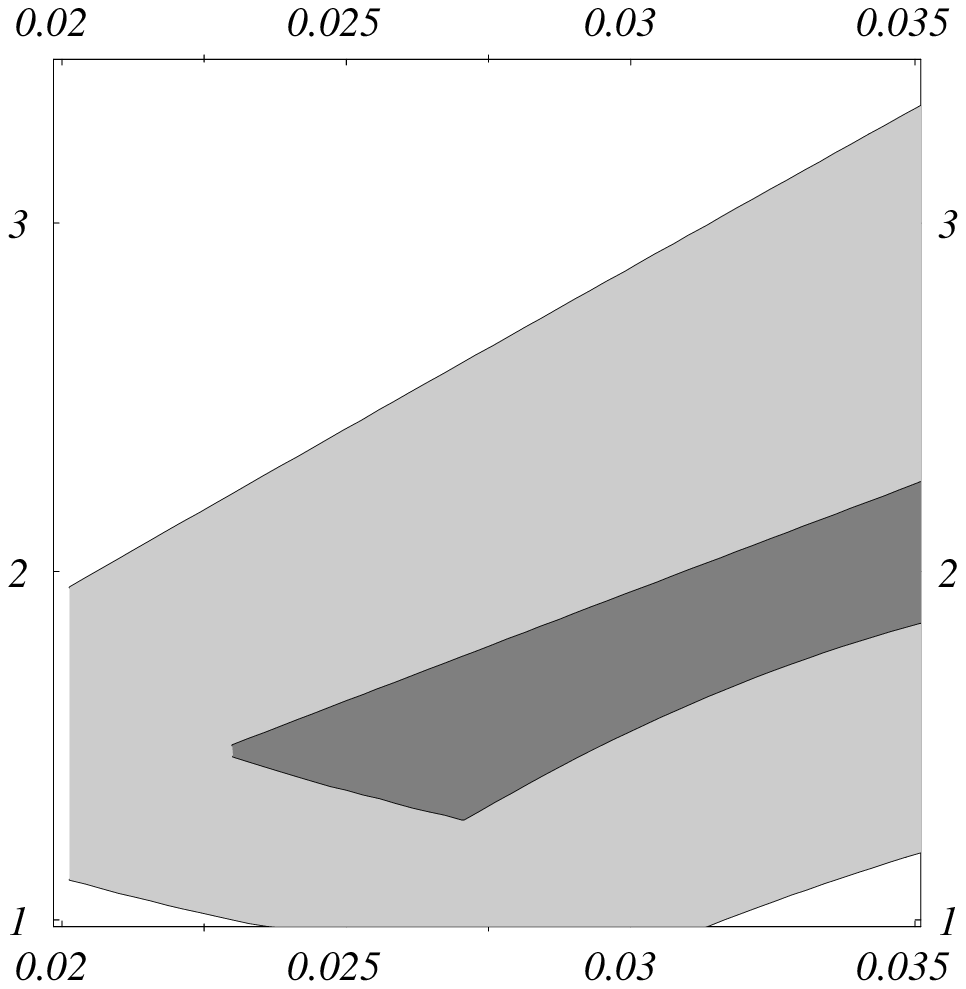}{Relation between $\BR(K_L \to \pi^0 \nu
  \bar\nu)$ and $R_{\Delta m_B}$ in the SM: A scan over the presently allowed
  region for ($\bar\rho,\bar\eta$) (c.f. Fig.~\ref{nARrhoeta}) yields
  the dark (light) region in the $R_{\Delta m_B} - \BR(K_L \to \pi^0
  \nu \bar\nu)$ plane for the central values (total intervals) of the
  parameters $\xi, X^2(x_t)$ and $A$.}
  {\widthAR}{\heightAR}{\spaAR}{R_{\Delta m_B}}{\BR(K_L \to \pi^0 \nu
  \bar\nu) \times 10^{11}}


\end{document}